\documentclass[twocolumn]{article}
\usepackage{graphicx}

\begin{document}
\title{A possible explanation for Earth's climatic changes in the
past few million years}
\author{W. W\"olfli \footnote{Institute for Particle Physics, ETHZ H\"onggerberg, CH-8093 Z\"urich, Switzerland; e-mail: woelfli@particle.phys.ethz.ch}\hspace{0.4em} and W. Baltensperger
\footnote{Centro Brasileiro de Pesquisas F\'\i sicas, r.Dr.Xavier Sigaud,150, Urca, 22290 Rio de Janeiro, RJ, Brazil; e-mail: baltens@CBPF.br}}

\date{\today}
\maketitle
\begin{abstract}\noindent
The astronomical theory of Milankovitch relates the changes of Earth's
past climate to variations in insolation caused by oscillations of the
orbital parameters. However, this theory has problems to account for
some major observed phenomena of the past few million years. Here, we
present an alternative explanation for these phenomena. It is based on
the idea that the solar system until quite recently contained an additional
massive object of planetary size. This object, called Z, is assumed to
have moved on a highly eccentric orbit bound to the sun. It influenced
Earth's climate through a gas cloud of evaporated material. Calculations
show that more than once during the last 3.2 Myr it even approached the
Earth close enough to provoke a significant shift of the geographic
position of the poles. The last of these shifts terminated the Earth's
Ice Age epoch. The origin and fate of Z is also discussed.
\end{abstract}
\tableofcontents
\section{The problem of Earth's past climate changes}
\begin{figure*}
\includegraphics[width=16cm,keepaspectratio]{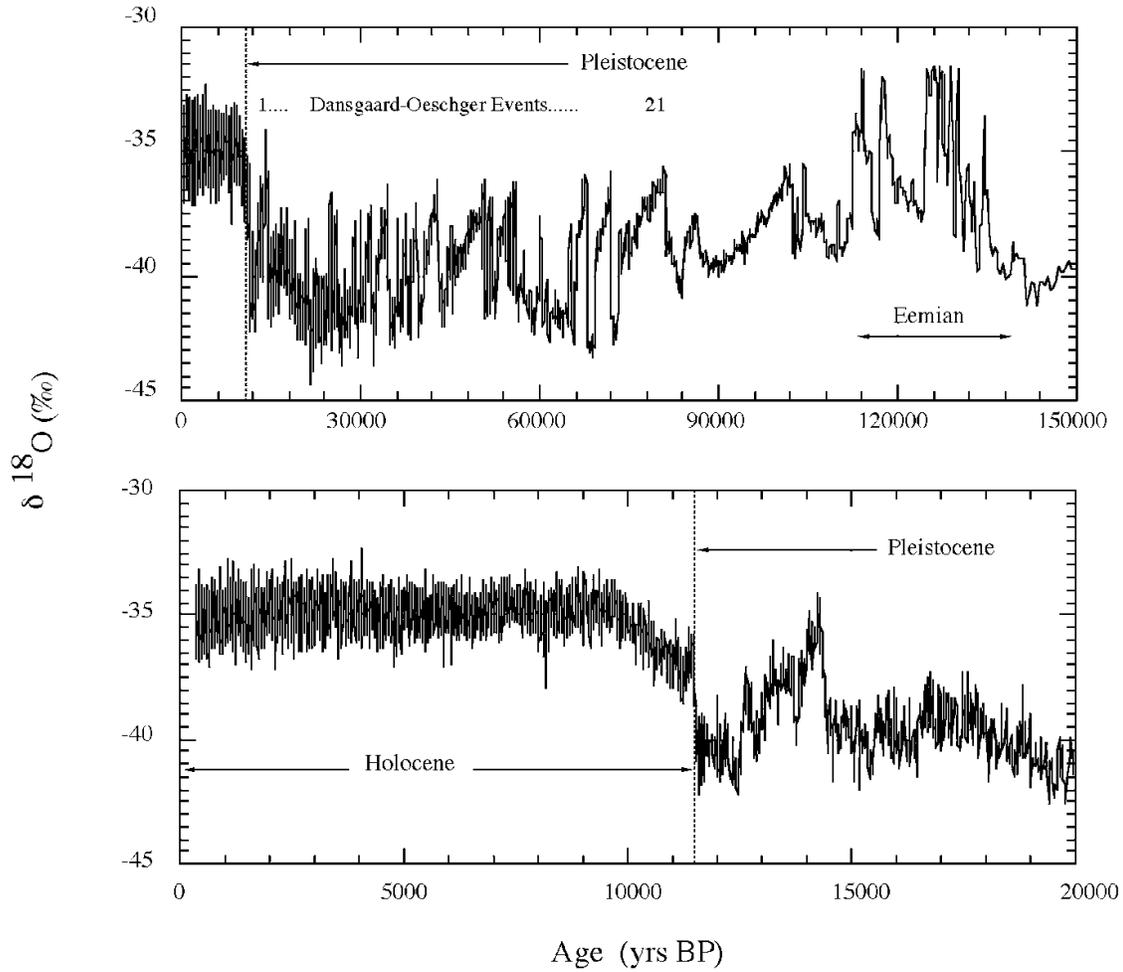}
\caption{Upper part: Ice core $\delta^{18}$O record (GRIP) over
150 kyr \protect\cite{Greenland}. This time span covers the last
Interstadial, the Eemian, the last glacial period (ca. 90 -- 11.5
kyrs) which is characterized by transient and fast temperature
variations (21 Dansgaard-Oeschger events), and the Holocene
reflecting the sudden improvement of Earth's climate at about 11.5
kyrs ago. The lower part is an expanded view of the same record
for the last 20 kyrs. It shows the time structure of the
Dansgaard-Oeschger event No.1 and the fast Pleistocence-Holocene
transition as indicated by the vertical dashed line.}
\label{IceEnd}
\end{figure*}

Ancient climate is recorded in deep sea sediments, polar ice and
some other natural archives mainly through changes in the oxygen
isotope ratio, the $\delta^{18}$O. In deep sea sediments the
$\delta^{18}$O-signal reflects the combined effect of water
temperature and of the amount of ice on Earth, whereas in polar
ice it is a reliable proxy for the ambient local surface air
temperature at the time of freezing. The study of paleontology
and iso\-to\-pic composition of hundreds of drilling cores
extracted from these archives during the last few decades confirm
that Earth's past climate changed between colder and warmer
phases, known as the glacial-interglacial cycles. Spectral
analysis of this kind of proxy data covering the Late Pleistocene
have yielded periodicities, which agree more or less with those of
Earth's main orbital parameters, i.e. precession (23 kyr),
obliquity (41 kyr) and eccentricity (100 kyr), which appear at
first sight to support the old idea of Milankovich
\cite{Milankovitch}, that Earth's climate reflects the variation
in insolation (solar heating) caused by the combined effect of the
variation of these parameters \cite{Imbrie93,Imbrie84}. However,
this astronomical theory of climate has several shortcomings. In
particular, the main glacial-interglacial transitions occurred
approximately every 100 kyr. This feature contradicts the
astronomical calculations which show that the modulation of the
insolation due to the eccentricity oscillation is negligibly small
compared to that of the two other orbital parameters. Many
hypotheses have been formulated to explain this ``100 kyr
problem'', sometimes involving additional astronomical parameters,
such as the oscillation of the inclination of Earth's orbital
plane, whose periodicity equals that of the eccentricity
\cite{Muller, Muller97}. Most of the more recent attempts to
understand this problem rely either on a non-linear response of
the ice-sheet dynamics \cite{Paillard}, or of that of the coupled
Atmosphere-Ocean system to the astronomical insolation forcing
\cite{Ganalpolski}. Although the results of some of these models
compare well with the time spectra of the climate record, up to
now none of them could reproduce the amplitude and phase of each
glacial cycle. For instance, one of the most prominent
interglacial periods of warmings observed around 400 kyr ago,
known as the ``isotope stage 11''- problem, happened at the time
when the solar insolation was smaller than during the glacial
times prior to and after this period. More recent measurements
with improved time resolutions, on deep sea sediment
\cite{Adkins,Tiedemann} and polar ice cores, GRIP
\cite{Greenland}, GISP2 \cite{Grootes} and Vostok \cite{Petit}
revealed some additional and most puzzling features of past
climate. Fig.~\ref{IceEnd} shows the well known
$\delta^{18}$O-record of the GRIP ice core covering the last 130
kyr of Earth's climate history, including the last interstadial,
the Eem, the last glacial maximum around about 18--20 kyr BP, and
its abrupt termination at about 11.5 kyr ago. Although some
discrepancies between the GRIP and GISP2 data older than about 100
kyr BP have been noted \cite{Grootes}, it is evident, that Earth's
climate changed rapidly during the last Ice Age period. These
temperature peaks, called Dansgaard-Oeschger events, cannot be
explained in terms of the Milankovitch effects, since they are too
fast in comparison with the orbital parameters induced solar
insolation variations. Even more puzzling is the fast termination
of the last Ice Age period, the Pleistocene-Holocene transition,
shown in the lower part of Fig.~\ref{IceEnd}. This transition
ended the Pleistocene, and the following Holocene is a period of
warmer temperature and generally more stable climate. It has been
demonstrated that the climatic system in the Northern hemisphere
is capable to switch rapidly from one given state to another
\cite{Stocker}. But again, if for the same reason as mentioned
above the Milankovitch effect cannot be the cause of such a
switch, what else could it be? Is it possible that the luminosity
of the sun varied in accordance with these observations? This is
very unlikely, because prior to the Ice Age epoch, i.e. during the
Late Pliocene until about 3.3 Myr ago, the climate was mild,
stable for millions of years, and very similar to what it has been
since the beginning of the Holocene \cite{Tiedemann}. Why should
the irradiance of the sun suddenly get noisier for a limited time?
The Milankovitch effect and small changes of the solar irradiance
are of relevance for Earth's climate. However, we agree with some
of the critics pointing out that both possibilities fail to
explain many properties of the observed past changes of climate
and that new ideas are required to solve the riddle of the Ice Age
\cite{Muller,Paillard}.

In what follows we present such a new idea. First, we show that the
fast termination of the last Ice Age period, the Pleistocene-Holocene
transition, can best be understood in terms of a rapid polar
wandering, as has been suggested (and refused) more than 100
years ago. We then demonstrate that such a shift can occur during
a close encounter between an object of planetary size and the Earth.
Assuming that this object, called Z, was on a highly eccentric orbit
bound to the sun, we find that it could be responsible for Earth's
climatic changes over the last few million years. Z must have
disappeared from the solar system some time after the polar shift
event, and we also consider the mechanisms responsible for its
disappearance.

\section{Evidences for a polar shift event at the end of the last Ice Age period}
The Last Glacial Maximum (LGM) around 18--20 kyr BP is characterized
by the lowest known temperature in Earth's history, low precipitation
rates, reduced sea level of at least 120 Meter \cite{Fairbanks},
and huge polar ice shields extending over the northern part of
America down to a latitude of almost 40$^\circ$ N during the last
glacial maximum (Wisconsin stage), and in Europe to about 50$^\circ$ N,
known as Weichsel and W\"urm stages in its northern part and in the
Alps, respectively \cite{Fullerton}. It is also well documented
that at the same time large flocks of Mammoths, Woolly Rhinos and
many other species of mammals roamed in north-eastern Siberia and
north-western Alaska \cite{Whitley}. Remains of these animals have been
found on the New Siberian Islands as far as 76$^\circ$ North
indicating that food was available to support these largest mammals
throughout the Late Pleistocene. It has also been noted, that at
the same time this part of the world was covered by non-arctic
forests up to about the same latitude, e.g. more than 1000 km
within present days' arctic cycle (66,66$^\circ$). Even more
puzzling, however, is the evidence that at the end of the last ice
age, the so called Pleistocene-Holocene transition, these animals
were not only instantaneously killed by some yet unknown cause,
but also immediately deep frozen together with the surrounding
mud \cite{Cuvier}. The fact that the flesh of some of these animals
was still eatable (for dogs) at the time of their discovery about
200 years ago suggests temperatures below freezing point throughout
the Holocene in this part of the world. This is in contrast to the
general warming up observed in many other places on Earth including
both polar regions. The observation of simultaneous mass extinctions
of small and large animals in the western part of Alaska, in Western
Europe and North America \cite{Scott,Martin} and the distinct sea
level rise leave no doubt that the observed fast rise in temperature
at the end of the last ice age period was accompanied by a global
cataclysm and a rapid reorganization of the global climate.

During the LGM the northern polar ice cap was not centered at the
position of the (present day) north pole, but displaced by about
18--20 degree southwards on the American side of the globe
\cite{Fullerton}. This introduced the possibility of a rapid
polar wandering, i.e.\ a change in orientation of the Earth with
respect to its rotation axis. This idea was considered in detail
more than 100 years ago by the most eminent scientists of that
time such as G.H. Darwin, J.C. Maxwell, G.V. Schiaparelli,
W. Thomson and many others. They concluded that such a turn of
the globe changing the geographical position of the North Pole
to the amount required by the observations is in principle possible,
but would require a force sufficient to overcome the stabilizing
effect of the equatorial bulge of our planet, which acts like the
rim of a gyroscope. Since they were unable to find any terrestrial
or extraterrestrial mechanism that could give rise to such a force,
this idea was finally dismissed as impossible and, in fact,
``not worth discussing'' \cite{Hapgood}. Only more recently was
this verdict seriously questioned when the evidence from rock
magnetism studies unmistakably suggested that the wandering of
the geographic poles was more rule than exception in Earth's
history \cite{Clegg,Gold}. However, this wandering is a slow
process, probably connected to the continental drifts first
predicted by Wegener \cite{Wegener}, and therefore it cannot
explain the rapid termination of the last ice age. In view of this
dilemma, Hapgood \cite{Hapgood} suggested an alternative possibility
based on the idea that the observed changes reflect the occasional
sliding of the whole rigid crust over the viscous, plastic or
possibly fluid inner layers of our Earth. He showed that the mass
of the Antarctic ice cap is not symmetrically distributed about
the rotation axis and believed that the resulting centrifugal
effect is sufficiently large to initiate such a slippage, provided
the ice cap can re-grow large enough after each slide. In his
foreword to Hapgood's book, A. Einstein cautiously commented this
idea as follows: ``The only doubtful assumption is that the Earth's
crust can be moved easily enough over the inner layers''. This is
indeed the crucial point of this idea which eventually has not been accepted.

Was this the end of the polar shift debate, although a fast polar
shift is still the most straightforward explanation for all observed
effects connected with the fast termination of the last Ice Age
period? In further investigating this question, we found that
one particular excitation mechanism, namely the deformation of
the Earth in a transient external gravitational field gradient,
can produce the required effect.
\section{The excitation of a polar shift during a close flyby of
a massive object}
Any redistribution of Earth's mass is accompanied by a wobble
and turn of the entire Earth relative to its rotation axis. This
movement depends on both the initial mass redistribution and
the rate at which the Earth's rotational bulge can readjust to
the changing position of the rotation axis on the globe. Generally,
the polar wandering resulting from Earth-bound causes are either
too slow (drift of the continental plates), or too small (storm,
ocean currents), to produce the rapid and large shift required to
explain the fast termination of the last ice age period. But what
about cosmic events? There are two possibilities worthwhile to be
considered: A collision of an object with the Earth, and a close
flyby of an object. The first case has previously been discussed
in the Literature \cite{Kelly}. It can be dismissed, because an
impact, capable to significantly alter Earth's angular momentum
or its moment of inertia, would have left one or several craters
and other distinct geophysical and chemical signatures, which would
have been discovered by geologists, bearing in mind that the event
we are discussing occurred only 11.5 kyr ago. We therefore consider
the possibility that the Earth had a close encounter with a massive
object. We estimate first the minimum amount of deformation required
to cause a polar wandering of about 18$^\circ$, as is mentioned in
chapter 2. We then show that such a shift can be induced by an object
of planetary size during a flyby and that it is fast enough to explain
the rapid termination of the last Ice Age epoch.

\subsection{Earth's deformation in a strong gravitational field gradient}
Prior to any disturbance the rotating Earth is to a large extent
in a hydrostatic equilibrium. The moment of inertia is a symmetric
tensor $\Xi$ which connects the vectors of angular
velocity $\vec\omega$ and angular momentum $\vec D$ by the
equation $\vec D = \Xi \vec\omega$. For the undisturbed Earth the
diagonalized tensor has the elements $\Xi [1,1]\equiv A = \Xi [2,2]
\equiv B = 8.010247\cdot 10^{37}$ kg m$^2$ and $\Xi [3,3]
\equiv C = 8.036559\cdot 10^{37}$ kg m$^2$ \cite{Melchior}.
The ratio $C/A = 1.0032847$ shows that the equatorial bulge
contributes about 3.3 \%$_0$ to the main principal moment of
inertia; the radius at the equator exceeds that at the poles by
about 21 km. This equatorial bulge stabilizes the position of the
rotation axis along the figure axis C. During the flyby of a massive
object, the globe is transiently deformed so that the new principal
axis of the inertial tensor $\Xi$ --- the figure axis --- deviates
from the rotation axis. These axes then precess around each other,
whereby the angular momentum $\vec D$ remains fixed in space. For
an observer on Earth the angular velocity vector $\vec \omega$ moves
around the figure axis. In co-ordinates fixed to the Earth this free
motion is described by the Euler equation
\begin{equation}\label{1}
\frac{d \Xi\, \vec\omega}{dt} = [\Xi\,\vec \omega ,\vec \omega]
\end{equation}
where the square bracket represents the vector product. According
to Eq.~\ref{1} the precession frequency of a rigid Earth would
be $\omega_0 (C-A)/A $, with $\omega_0$, the angular velocity of
Earth's rotation. This yields an ``Euler period'' of 305 d. A minute
precession of the rotation axis was observed by S.C. Chandler more
than 100 years ago, but its period turned out to be 435 d \cite{Chandler}.
Newcomb \cite{Newcomb} showed that this discrepancy can be understood
by assuming that the Earth is not rigid but can be plastically
deformed, which was quite a pioneering conclusion in those days.
Since then, this so called Chandler's free motion of the rotation
axis has been studied in great detail, but always in view of small
deviations only \cite{Munk,Wahr}. The presently continuing Chandler
motion is still open to discussion \cite{Chao}.

A massive object passing the Earth at a close distance deforms it
with a time dependent tidal force, which is the difference between
the gravitational force at the given point and that acting at the
Earth's center. When the distance $R$ between the centers of the two
masses is large compared to the radius $R_E$ of the Earth, the tidal
force $\vec F(z)$ is parallel to the z-direction which points to the
perturbing mass $M_Z$. It has the value
\begin{equation} \label{2}
F(z) = 2 M_Z G z / R^3
\end{equation}
where $G = 6.673\cdot 10^{-11}$ m$^3$kg$^{-1}$s$^{-2}$ is the
gravitational constant. On Earth's surface $z = R_E \cos \gamma$,
where the angle $\gamma$ is the latitude. This tidal force increases
as the third reciprocal power of the distance $R$ between the
center of the perturbing mass and that of the Earth. For example,
if the moon would be 10 times closer to the Earth, then its tidal
force would be 1000 times larger than it is at present.

In first order approximation, this force tends to give the Earth an
ellipsoidal shape. Such a deformation is possible because Earth's
interior is liquid or plastic except for the inner core and the
peripheral crust \cite{Melchior}. We estimate the deformation due
to the force, Eq.~\ref{2}, in a static approach by minimizing the
energy as a function of the deformation in the external field. For
this we assume that the redistribution of material at the surface of
the Earth is given by
\begin{equation}\label{3}
H(\gamma ) = H_0 (\cos (2 \gamma ) + 1/3)
\end{equation}
where $H_0$ is the amplitude of the tide. The energy of the material
in the deformed state minus that prior to the deformation in the
presence of the perturbation is then
\begin{equation}\label{4}
\Delta E(H_0) =\frac{32 \pi}{45} \rho R_E^2 \left( H_0^2 g -
\frac{H_0 M_Z G R_E^2}{R^3}\right)
\end{equation}
which is minimized with
\begin{equation}\label{5}
        H_0 = R_E^2 M_Z G / (2 R^3 g)
\end{equation}
Here $g$ = 9.8 m/s$^2$ is the gravitational acceleration at the
surface of the Earth and $\rho$ the specific weight of the displaced
material (3500 kg/m$^3$ for the Earth's surface). Applying
Eq.~\ref{5} to the Earth-Moon system we find that the peak-to-peak
level difference on Earth induced by our Moon is $2 H_0$ = 0.35 m
which agrees quite well with the recently evaluated deformation of
about 0.30 m \cite{Geiger}. We use this formula to estimate the
deformation produced by a flyby of a massive object, although we
are aware that this simple approach probably overestimates the
deformation remaining immediately after such a flyby. In particular,
if the response to the strong tidal force is too fast, then part of
the deformation will quickly disappear when the force ceases. We
also expect that the actual deformation will not be a linear function
of the gravitational force, and will depend on the relative motion
of the perturbing mass to the surface of the rotating Earth.

To estimate the minimum amplitude $H_0$ of the tidal bulge required
to cause a polar wandering of about 18$^\circ$, as explained in
chapter 2, we added to the equilibrium deformation of the Earth an
additional mass distribution of the form of Eq.~\ref{3}, which was
rotated by an angle $\alpha$ with respect to the original figure axis.
The diagonalization of the new inertial tensors resulted in a new
figure axis whose tilting angle relative to the former axis depends
on $H_0$ and $\alpha$. Minimizing $H_0$ as a function of $\alpha$,
we find that a tilting angle of 18$^\circ$ can be achieved with a
bulge amplitude of $H_0$ = 6.45 km and an angle $\alpha$ = 30$^\circ$.
This result allows us to estimate the mass of our flyby-object or
its distance of closest approach, by means of Eq.~\ref{5} which
fixes $M_Z/R^3$. Assuming for instance that the distance of closest
approach should be larger than the Roche limit, which for our Earth
is about $R_l = 2.44 R_E$, i.e. 15'600 km (assuming equal density of
both objects), then the mass $M_Z$ must be larger than about 0.03 times
the mass of the Earth ($M_E$), a value which is somewhat smaller than
that of Mercury. On the other hand, 2-body dynamics tells us that
the mass has to be much smaller than that of our Earth. Otherwise
Earth's orbit, in particular its inclination and eccentricity, would
have been altered during such an event to an amount which would be
incompatible with the present orbital parameters of the Earth. This
consideration suggests an upper limit of about $M_Z < 0.2 M_E$, and,
according to Eq.~\ref{5}, a maximum distance of closest approach of
29\thinspace 300 km. Since our static approximation most probably
overestimates the tidal effects, we suspect that the mass of Z was
larger than 0.03 $M_E$ and that its closest approach during this event
was even within the Roche limit.
\subsection{The fast polar wandering}
\begin{figure*}
\includegraphics[width=15.9cm,keepaspectratio]{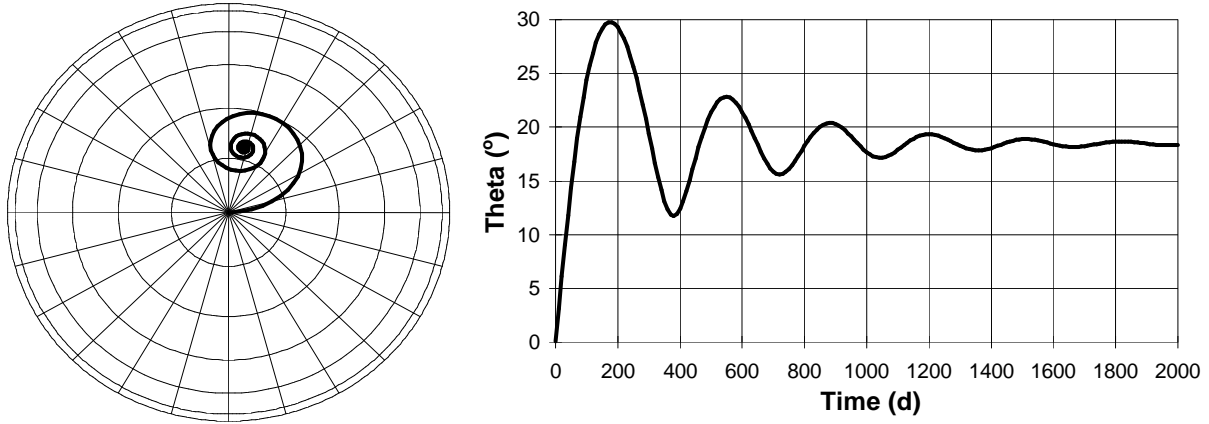}
\caption{Left side: Geographic movement of the rotation axis
induced by the flyby of a massive object, assuming the deformation
and relaxation parameters discussed in 3.1. and 3.2. The circles
are latitudes with respect to the previous pole position, spaced
15$^\circ$. Right side:
Time dependence of the angle $\Theta$ between the angular momentum
(fixed in space) and the previous pole direction.} \label{PShift}
\end{figure*}

The sudden deformation of the Earth resulting from a short flyby event
(the excitation time is less than 10 h) defines the new figure axis
around which the rotation axis immediately starts to precess, as
described by Euler's Eq.~\ref{1}. Since the figure axis itself moves
only by small amounts in co-ordinates fixed to the Earth, the rotation
axis effectively spirals to the shifted position of the figure axis.
To demonstrate this polar wandering, we solved Eq.~\ref{1} numerically.
Since the Earth is not a rigid body we started the calculation with the
inertial tensor, as defined by the initial deformation, and assumed
that it relaxes towards the equilibrium tensor $\Xi_0$ oriented along
the instantaneous rotation axis as
\begin{equation}
\frac{d\Xi}{dt} =-\frac{\Xi (t) -\Xi_0 [\vec \omega (t)]}{\tau}
\end{equation}
Thus we approximated the relaxation behavior by a single
relaxation time $\tau$, which includes both the capability of the
equatorial bulge to follow the rotation axis and the disappearance
of the deformation responsible for the polar wandering. Assuming
a relaxation time of $\tau$ = 500 d, we find that the rotation
axis moves into the new position with a precession period of about
400 d. The final destination is reached within about four precession
periods or about 4.8 years (Fig.~\ref{PShift}).

This is certainly fast enough to explain the rapid termination of
the last Ice Age period shown in the lower part of Fig.~\ref{IceEnd}.
We have to add here, that the true relaxation behavior of the Earth
in the presence of a large, but short-term deformation is still open
to speculation. The theoretical estimates for this relaxation time
vary between a few 10 d and 10 yr \cite{Smith}. We find that $\tau$ in
our example has to be larger than about 100 days to achieve a polar
shift of the required magnitude. \subsection{The impact of the polar
shift event on the climate}
The lower part of Fig.~\ref{IceEnd} shows, that the termination of
the last ice age period is characterized by two events: A short term
increase of temperature around about 14.5 kyr. ago, which was followed
by a slow decrease back to the Ice Age level. The following cold
period, known as Younger Dryas (YD), lasted for about 800 years.
At about 11.5 kyr ago the temperature suddenly rose again, but, and
in contrast to the previous event, it continued to increase slowly
until about 9000 years ago. From then on the surface temperature
over Greenland essentially remained at that high value up to the
present day. Can we understand why the first temperature increase
was a transient episode only, whereas the second event had such a
long lasting effect? To answer the first part of this question we
need to consider an additional property of our flyby object, which
will be discussed in the next chapter. It is sufficient to mention
here that in this case the flyby distance of our object was simply
too large to excite a significant polar wandering. The consequences
of the second event can be explained in terms of the polar wandering
discussed above. For this we have to remember (see chapter 2) that
the shape and extension of the ice shield in the Northern hemisphere
prior to the shift suggest a pole position at a longitude of about
90$^\circ$ W and a latitude of about 72$^\circ$ N, which is quite
close to the present position of the North magnetic pole. Identifying
this position with the starting point of our calculation presented
in Fig.~\ref{PShift}, we see that the net effect of this polar
wandering was a rotation of the globe by about 18$^\circ$. As a
consequence, the latitudinal pattern of the incident solar radiation
immediately changed, so that, e.g. the low latitude part of the North
American and European ice shields rapidly melted away, whereas in
formerly ice free regions of the Arctic Ocean permanent ice was
formed. Similarly, in the eastern part of the Antarctic the extent
of the ice shelf was reduced, and, as we suspect, the formerly ice
free western coast was covered to the extend observed today. After
a transition time of about 2\thinspace 000 years during which the
polar ice sheets adjusted to the new pole position, the new state
of the climate, called Holocene, was established. Compared to the
previous climate state, the Holocene is significantly warmer,
essentially, because the North pole moved away from the upper part
of the North American continent into the present position in the
Arctic Ocean. This resulted in an increased heat exchange between
ice and water, which significantly reduced the extent of the polar
ice cap. The albedo decreased accordingly. This reduction was only
partially compensated by the corresponding changes in the southern
hemisphere. According to this scenario, the Holocene is a new stable
epoch and not just another Interstadial, as is generally believed,
because the new pole position is stable and Z no longer influences
the Earth.

\section{Planet Z: a transient member of the solar system}
We have seen that an object of mass $M_Z$ between 0.03 $M_E$ and
0.2 $M_E$ can excite a polar shift which is both fast and large
enough to explain the termination of the last Ice Age, provided
it passed the Earth at a distance of less than about
30\thinspace 000 km. Since it is extremely unlikely that an
object on a hyperbolic trajectory would accidentally approach
the Earth so close during one singular flyby, we consider the
possibility that it was moving on a bound orbit extending beyond
that of the Earth. The problem with this assumption is that this
object, Z, no longer exists --- otherwise we would observe it --- so
that the occurrence of the proposed close encounter at a specific
time cannot be verified by backward calculations. Therefore, we
shall consider the long term behaviour of Z on the basis of assumed
orbital parameters in order to find out whether close approaches
to the Earth occur as often as expected. We shall see that this
``trial and error'' approach not only gives a positive answer, but
even provides an explanation for the frequent fast climatic changes
observed prior to the pole shift event (see chapter 5).

\subsection{The orbit of planet Z and its close encounters with the Earth}
There are constraints on the initial orbital parameters of Z.
Its aphelion has to be larger than that of the Earth, otherwise
it could not approach the Earth close enough. We shall see further
below that at each perihelion passage a surface layer of Z has to
evaporate at sufficiently high temperatures so that atoms reach
the escape velocity. This requires an extremely eccentric orbit
with $a \ll b$, where $a$ is the perihelion and $b$ the aphelion
distance. Thus the major semi-axis $A = \frac{1}{2}(a +b) \approx
\frac{1}{2}b$ must be larger than 0.5 AU, and the eccentricity
$\epsilon = (1 - a/A)$ close to one. Expecting that the encounter
frequency rapidly decreases with increasing $A$, we restrict the
investigations, to the range $0.5 < A < 2$ AU. There are no
restrictions on the inclination $i$ of Z's orbit, so that the full
range $0^\circ < i < 180^\circ$ has to be explored.

With this in mind, Nufer et al.~\cite{Nufer} calculated the movement
of Z in the gravitational field of the sun, the Earth, the Moon and
all other planets, except Uranus, Neptune and Pluto, and evaluated
by means of a Pascal program its closest approaches to the Earth
for each orbital period. Details of this program, its performance,
and the relevant results shall be published in \cite{Nufer}. In
the simulations Z was always added to the solar system at the year
J2000.0, ignoring the fact that at this time it no longer existed.
So far calculations have been performed, forward as well as
backwards in time, for two different masses ($M_Z = 0.11 M_E$ equal
to that of Mars, and $M_Z = 0.06 M_E$) and seven different sets of
initial orbital parameters ($A = 0.978$ and 2 AU, six different
inclinations in the range $i = 0^\circ$--$180^\circ$ and two
eccentricities $\epsilon = 0,973$ and 0.987) The time ranges varied
between 150 and 750 kyr. As expected, we found that encounters are
more frequent for $A = 0.978$ than for $A = 2$. Their number also
increases with increasing eccentricity $\epsilon$, and is largest
for small inclinations $ i< 6^\circ$. The inclination is defined
here as the angle between Z's orbital plane and the invariable
plane which is perpendicular to the orbital angular momentum of
the solar system. The results obtained with the initial conditions
$A = 0.978$, $\epsilon = 0.973$, $i = 0^\circ$ and $180^\circ$,
respectively, are too similar to decide whether Z was on a pro-
or a retrograde orbit. Since the computing speed strongly depends
on the precision assumed for the numerical integration procedure,
we restricted it to a moderate value of $10^{-13}$ per integration
step, which was sufficient for most considerations. With this
restriction, about 15 h were needed to cover 10 kyr using a 133 MHz PC.

\begin{figure*}
\includegraphics[width=16cm,keepaspectratio]{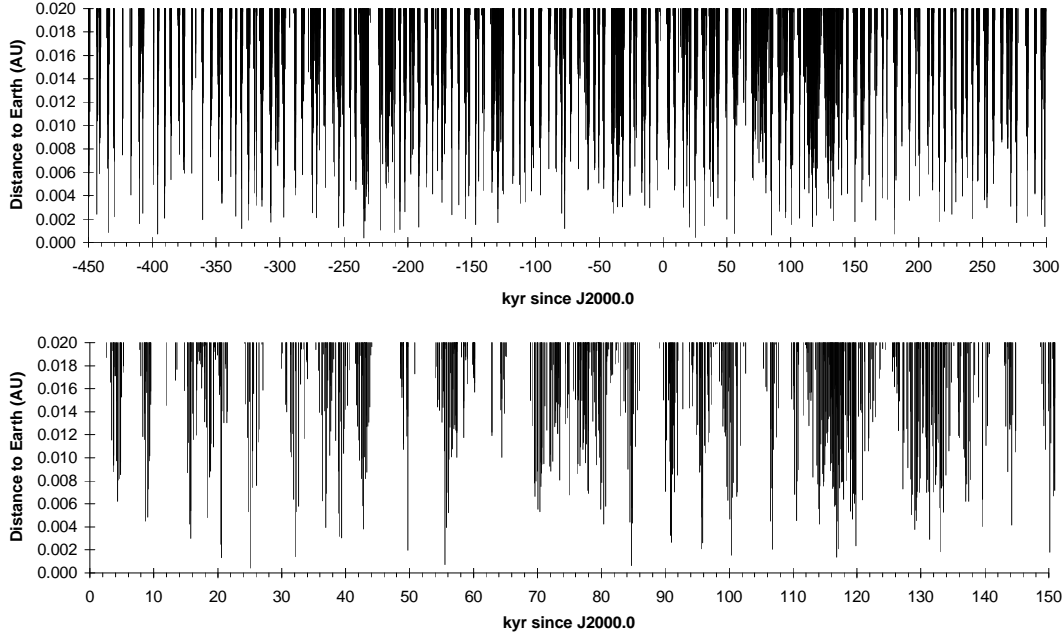}
\caption{Upper panel: Closest approaches of planet Z ($M_Z = 0.11
M_E$) to the Earth over 750 kyr and to distances of less than 0.02
AU = $3\cdot 10^6$ km as indicated by the endpoint of each
vertical line. Lower panel: Expanded view of the irregular
clustering of these approaches over 150 kyr.} \label{MoveZ}
\end{figure*}

\begin{figure*}
\includegraphics[width=16cm,keepaspectratio]{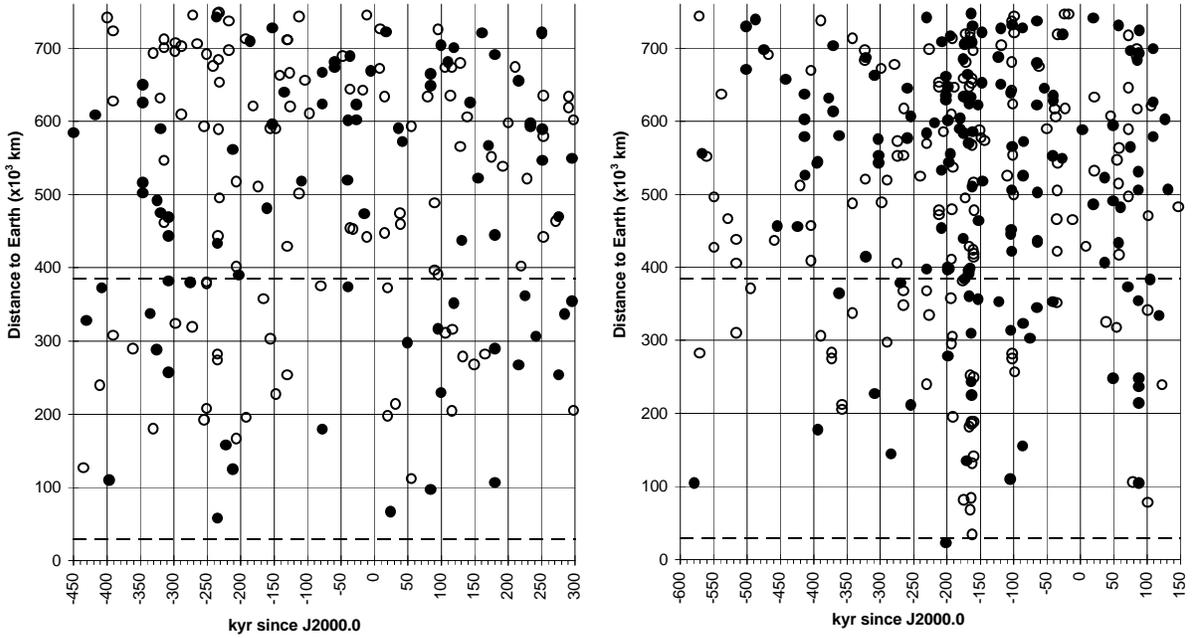}
\caption{Closest approaches over 750 kyr below twice the Moon-Earth
distance. The pattern depends on the selected orbital parameters
and the mass of the object. The left side shows the result obtained
for a mars-like object ($M_Z = 0.11 M_E$), and the right side that
for an object with $M_Z = 0.06 M_E$ but otherwise identical start
parameters. The two horizontal dashed lines in both figures indicate
Moon's distance (384\thinspace 000 km), and the critical distance
below which significant polar shifts have to be expected
(30\thinspace 000 km), respectively. The open (filled) circles
mark the encounters of the objects on their way towards (from) the sun.}
\label{flyby}
\end{figure*}

The result of one of these calculations is displayed in Fig.~\ref{MoveZ}.
It shows the closest approaches of Z ($M_Z = 0.11 M_E$) to the
Earth over a time period of 750 kyr. The initial parameters at
J2000.0 were $A = 0.978$ AU, $\epsilon = 0.973$ and $i = 0^\circ$.
Plotted are all encounters with distances of less than 0.02 AU
= $3\cdot 10^6$ km. The irregular clustering of these approaches
is the result of the osculating orbital parameters of Z and the
Earth. An expanded view of all closest approaches with distances
less than twice that of the Moon over 750 kyr is given in Fig.~\ref{flyby}.
To illustrate the mass dependence we also show the pattern obtained
for an object with $M_Z = 0.06 M_E$, but otherwise identical
orbital parameters. The two horizontal dashed lines in both
figures indicate Moon's distance (384\thinspace 000 km), and
the critical distance below which significant polar shift events
have to be expected (30\thinspace 000 km). In the first case we
note a closest approach of about 59\thinspace 000 km, i.e. no pole
shift event, but a distinct one in the second case (23\thinspace 500
km). From these and the other trials we conclude that an object
moving on a highly eccentric orbit approaches the Earth to polar
shift distance about once in 1 Million years, which is compatible
with the duration of the ice age epoch (see chapter 6.1). It is
meaningless that the calculated polar shift event does not occur
at the expected time in view of the fact that the exact mass and
orbit of Z is unknown. In both cases we also observe about 6 flyby's
per 100 kyr below 300\thinspace 000 km. In all these events the
gravitational interaction is still strong enough to excite strong
earthquakes and volcanic activities (see chapter 7). All these close
flyby's also influence Moon's orbit. We found, however, that the
resulting disturbances over the investigated time ranges are averaged
to the amount compatible with the present orbit \cite{Nufer}. The
frequency of the flyby's increases with their distance. For distances
larger than twice that of the Moon they are harmless regarding tidal
effects induced on Earth, but they can still influence Earth's climate.
To understand this additional effect, which is further discussed
in chapter 5, we next examine what happened to Z on the proposed
highly eccentric orbit. For convenience, we assume that it was a
Mars sized object with $M_Z = 0.11 M_E = 6.58\cdot 10^{23}$ kg,
and radius $R_Z = 3.39\cdot 10^6$ m.

\subsection{Solar radiation effects: the gas cloud}
An object moving on a highly eccentric orbit is exposed to intense
solar radiation. The total amount of energy $E_{Zs}$ received by Z
per unit of surface and orbital period is
\begin{equation}\label{6}
         E_{Zs} = P_s\oint \frac{dt}{4\pi R^2}= \frac{M_Z}{2 L}P_s
\end{equation}
where $P_s = 3.73\cdot 10^{26}$ J/s is the total power radiated by
the sun. The orbital angular momentum $L = M_ZR^2 \frac{d\varphi}{dt}\quad (M_Z$:
mass of Z, $R$ its distance to the sun, and $\varphi$ the angle
co-ordinate) is determined by the major semi-axis $A$ and the
eccentricity $\epsilon$
\begin{equation} \label{7}
                L  = M_Z \sqrt{M_S GA(1 - \epsilon^2)}
\end{equation}
$G$ is the gravitation constant, and $M_S$ the mass of the sun.
Therefore
\begin{equation}\label{8}
 E_{Zs} = \frac{1}{2}\frac{P_s}{\sqrt{M_SGA(1 - \epsilon^2)}}
\end{equation}
Using the same orbital parameters as in Fig.~\ref{MoveZ} ($A = 1$ AU,
$\epsilon = 0.973$), this formula yields an energy per unit surface
of $1.67\cdot 10^{11}$ J/m$^2$ per orbit. During each perihelion
passage the power density reaches a maximum of about 2.83
MW/m$^2$. This is sufficient to evaporate any type of rock on
the exposed side and also to heat up the resulting gas so that
its light components, mostly atoms, even reach the escape velocity
$v_e$, which is 5020 m/s for a mars-like object. If all the
radiation energy were converted into kinetic energy, then the
total mass per m$^2$ being accelerated to escape velocity would
amount to
\begin{equation} \label{9}
                            m_g = 2 \frac{E_{Zs}}{v_e^2}
\end{equation}
or $m_g$ = 13\thinspace 300 kg/m$^2$. This corresponds to an
ablation rate of almost 5 m per orbital period, assuming that
Z was composed of silicates or basalts with typical densities
of about 2700 kg/m$^3$. Since the conversion efficiency was hardly
100\% , this represents an upper limit for the total mass of
evaporated material. It amounts to $M_g = \pi R_Z^2 m_g = 4.8
\cdot 10^5$ Gt per orbital period, indicating that Z was surrounded
by a rapidly expanding gas cloud, with Oxigen and Silicon (atomic
weights 16 and 28, respectively) as its main components. Its
radius at the distance of Earth's orbit was $R_g = \tau_g v_g$,
where $\tau_g$ is the travelling time from the sun to Earth's orbit,
and $v_g$ the expansion velocity of the gas cloud. The temperature
$T$ of the atmosphere of Z from which the fastest atoms escape equals
the temperature of the escaped gas cloud. A measure of the radial
expansion velocity is $\sqrt{<v_g^2>} = \sqrt{\frac{k_B T}{m_a}}$,
with $k_B$ the Boltzmann constant, and $m_a$ the atomic mass of the
gas compound. Assuming a temperature of $T = 2000$ K during the
perihelion passage, this yields a velocity of 1000 m/s for Oxygen
atoms. In $\tau_g=34$ days, the travelling time of Z from the
perihelion to Earth's orbit, the gas cloud expanded to a radius
of 3 Mio.~km. The cloud remained spherical during this time, provided
the constituents of the cloud have no resonant transitions in the
solar spectrum. In this case, the Rayleigh scattering cross section
is of the order of the square of the classical electron radius, so
that less than one photon per atom was scattered during the travelling
time. In this case the light pressure effect is negligibly small,
and the scattered light is dominantly blue in colour. This behaviour
differs from that of an opaque grain which is accelerated inversely
proportional to its radius by the light pressure. A particle of
1 micron is displaced by many Mio.~km during 34 d.

It is interesting to compare these considerations with observations:
For instance, the coma of the comet observed in 1811 AD had a diameter
of $1.8\cdot10^6$ km \cite{Flammarion}, comparable to the estimate
obtained for Z. Its tail had a length of $179\cdot 10^6$ km and was
``white'' in colour suggesting that it contained dust grains of
various sizes and molecules with strong transition in the solar
spectrum. The appearance of Z was different. Its nucleus was much
larger than those of any presently known ordinary comets and it was
surrounded by a huge but essentially spherical gas cloud with bluish colour.

\subsection{Tidal effects on Z}
Z was also exposed to strong time dependent gravitational forces
during each perihelion passage. They induced a tidal wave, whose
kinetic energy was converted into heat so that its temperature
increased stepwise. To get an idea of this heating effect, we estimate
first the amplitude of this tidal wave. For this we use Eq.~\ref{5},
but this time with $M_Z$ for the solar mass, and $R_Z$ and $g_Z$ for
the radius and surface acceleration of Z ($M_Z$, $R_Z$, and $g_Z$ equal
to that of Mars). For a perihelion distance $R=a = 4\cdot 10^9$ m,
as determined by the orbital parameters $A$ = 1 AU and $\epsilon$ = 0.973,
we find a value of $H_0 = 3230$ m. The lowering of the energy through
the deformation in the tidal field, given by Eqs.~5 and 4, expressed
in variables for Z, is
\begin{equation}
\Delta E = - \frac{2}{15}\frac{R_Z^5GM_S^2}{R^6}
\end{equation}
where $M_S$ is the solar mass, and $R$ the distance to the sun.
During an orbit of Z twice this energy is transformed into heat,
i.e. $2\vert\Delta E_{Zt}\vert = 1\cdot 10^{25}$ Joule for $R=a$.
Tidal forces act on the sun as well. The corresponding values are
0.6 m and $-1\cdot 10^{23}$ Joule, respectively, smaller than the
tidal effect on planet Z. Comparing $\Delta E_{Zt}$ with the
influx of solar radiation which according to Eq.~\ref{8} is
$E_{Zs} \pi R_Z^2 = 6.05\cdot 10^{24}$ Joule; hence the values
are of the same order. Radiative energy primarily heated up the
surface, while the tidal deformation extended into the interior
of planet Z. Assuming that all of this energy was converted into
heat, then the temperature increase was $\Delta T = 2
\frac{\Delta E_{Zt}}{C_p M_Z} = 0.02$ K per orbit, assuming a
specific heat of $C_p = 650$ J/kg K. After about one Million orbits,
approximately equal to Earth years, the interior was melted and,
according to the Stefan-Boltzman $T^4$-law, tidal energy alone
could maintain a temperature of about 450 K on a black surface.
This implies that the surface was only molten during each
perihelion passage, whereas on its journey towards the aphelion
a thin solid crust was formed on the surface. Once liquefied,
Z easily reached its optimum shape during each perihelion passage.
This redistribution of matter was a turbulent process and occurred
twice during each passage. For smaller perihelion distances tidal
heating would dominate because of its $a^{-6}$ dependence per
orbit as compared to the $a^{-1/2}$ dependence of the radiation
energy given by Eq.~\ref{8}.

Tidal losses imply that the deformation lags behind the exciting
tidal field, so that the forces on Z are not symmetric on incoming
and outgoing positions. As a consequence, both the orbital energy
$E_0$ and the orbital angular momentum $L$ decrease. With decreasing
$E_0$ the major semi-axis $A$ of the orbit is reduced. The effect
of the tidal energy dissipation can be estimated by comparing the
maximum tidal loss $2 \Delta E_{Zt}$ with the orbital energy
determined by $E_0 = - \frac{M_Z M_S G}{2 A} = -2.65\cdot 10^{32}$ Joule.
The ratio $\frac{ E_0}{2 \Delta E_{Zt}} = 4.4 \cdot 10^7$ is a measure
for the number of orbits required for a substantial change of $A$.
The influence on the eccentricity $\epsilon$ is less evident. The
relation $\epsilon^2 = 1 + \frac{2 E_0 L^2}{M_Z^3 M_S^2 G^2}$ shows
that a decrease in $E_0$ lowers $\epsilon$, while a decrease
in $L$ increases $\epsilon$, since $E_0$ is negative. In view of
our restricted knowledge of Z, we are not able to state which of the
two effects prevailed.

\section{The climate impact of the gas cloud of planet Z}
The upper panel of Fig.~\ref{MoveZ} indicates that Z frequently
approached the Earth to distances within its surrounding gas cloud,
which had a radius of about 3 Mio.~km, according to our estimate.
We shall show that the incidence of this gas on Earth's atmosphere
resulted in a Greenhouse effect which was large enough to enhance
the global temperature. Using a simple model for this Greenhouse
heating we further demonstrate that the frequency structure of this
effect compares surprisingly well with that of Dansgaard-Oeschger
events observed during the Late Pleistocene (Fig.~\ref{IceEnd}).

\subsection{A gas cloud driven Greenhouse effect}
Each time the Earth passed through the gas cloud of Z, a fraction of
this material was captured by Earth's upper atmosphere. For a spherical
gas cloud, and constant gas density, the amount of gas $m_c$ captured
during one crossing is proportional to the ratio of the volume covered
by the Earth during its passage, $V_c \leq \pi R_E^2 2 R_g$, and the
volume of the gas cloud $V_g = 4 \pi R_g^3/3$. Neglecting the
gravitational focusing, which enlarges the capture radius $R_c$ over
Earth radius $R_E$ by at most 10 \% , the total mass of captured
particles is determined by:
\begin{equation}
                       m_c \leq 1.5 m_g R_E^2/R_g^2
\label{10}
\end{equation}
With the estimated mass for the gas cloud $m_g = 4.8\cdot 10^5$ Gt,
its radius $R_g = 3\cdot 10^6$ km, and Earth's radius $R_E = 6.4\cdot 10^3$ km,
this yields an amount of $m_c \leq 3$ Gt per encounter. The atoms
of this cloud hit the upper atmosphere of the Earth with a relative
velocity of about $v_r = 42$ km/s, assuming a crossing angle of
90$^\circ$ and equal velocities of about 30 km/s for the Earth and Z.
The particles are stopped at an altitude of about 150--200 km above
Earth's surface. Including the acceleration in the gravitational
field of the Earth (escape velocity $v_{es} = 11$ km/s), we find
that the total energy deposited at this altitude amounts
to $E_g = m_c (v_r^2+ v_{es}^2) /2 = 3\cdot 10^{21}$ J. Since the
passage through the cloud lasted up to 37 hours, this energy was
deposited into a thin layer of the uppermost atmosphere of the whole globe.

The impact of this cloud on Earth's atmosphere was manifold
considering the fact that each Oxygen atom had a kinetic energy
of about 160 eV. This energy was converted into radiation and
used to ionise and dissociate the main constituents of the atmosphere,
the Nitrogen and Oxygen molecules. The resulting ions and atoms
initiated a complex chain of chemical reactions in the upper atmosphere
during which the well known infrared-active compounds such as Ozone
(O$_3$) and Nitrogen oxydes (NO$_x$) were formed. By means of the
following simple arguments we can get some idea of the total amount
of infrared active molecules produced during such an event. Assuming
that 30 \% of the particle energy, or 50 eV, is consumed by dissociation,
then about 20 N$_2$ or O$_2$ molecules are decomposed per incoming
particle. The resulting 40 excited and ionised atoms interact with the
molecules of the atmosphere. Assuming that 50 \% of these atoms end
up in an infrared active molecule, then about 20 of these would be
produced per incoming particle. Since their mean molecular weight is
about twice the atomic weight of Oxygen, we expect that up to $20\cdot 2\cdot 3$
Gt = 120 Gt of Greenhouse gases, mostly Nitrogen compounds and Ozone,
could be produced during a flyby. True, this is less than the atmospheric
inventory of CO$_2$ which was of the order of 430 Gt during the last
glacial maximum. However, this deficit is more than compensated by
the fact, that the radiative
forcing of most of the infrared active molecules of interest here is
30--70 times larger than that of CO$_2$ for equal concentrations. We
conclude that the estimated amount of Greenhouse gases produced is
sufficient to transiently enhance the mean global temperature to a
level higher than what is predicted for a doubling of the CO$_2$
concentration \cite{IPCC}. The role of CO$_2$ and CH$_4$ during such
an event is open to discussion. According to the measurements made on
air trapped in the polar ice cores, their concentrations essentially
seems to follow the observed global temperature changes \cite{Barnola},
suggesting that they play a passive role only.

\subsection{The climate changes during the Ice Age epoch}
In chapter 3 we have explained why the climate was generally colder
during the Late Pleistocene than in the following Holocene. Here,
we show that the sudden increases in temperature of variable length
and frequency, the Dansgaard-Oeschger events shown in Fig.~\ref{IceEnd},
can be explained in terms of the Greenhouse heating effect discussed
above. For this we assume that the global temperature
increase $\Delta T$ is proportional to the Greenhouse gas
concentration, and that this value depends linearly on the number
of particles captured during each encounter. We further assume that
the suddenly enhanced gas concentrations decrease exponentially in
time after each encounter. In this case, the time dependent
temperature variations are given by
\begin{equation}
\Delta T(t) = T_0 \sum_j^{t_j<t}\sqrt{1 -\frac{D_j^2}{R_g^2}}\;
{\rm e}^{-\frac{t - t_j}{\tau_r}}
\label{11}
\end{equation}
were $D_j$ is the distance of closest approach of the flyby object
at time $t_j$, $R_g$ the radius of the gas cloud, and $\tau_r$ an
averaged residence time of the relevant Greenhouse gases. $T_0$ is
the initial temperature increase for $D_j \ll R_g$, which depends
on the amount of material captured from the gas cloud, the efficiency
of conversion into Greenhouse gases, and the radiative forcing of these.
The value of $T_0$ is open to discussion. For the reasons explained
above we are optimistic that $T_0$ is large enough to account for
the $\Delta T$ observed in the polar ice cores. We also note that
the residence times of the various Greenhouse gases of interest here
are not known. This crude model ignores long term reactions of Earth's
climate system to the sudden changes in the radiation budget.

\begin{figure*}
\includegraphics[width=16cm,keepaspectratio]{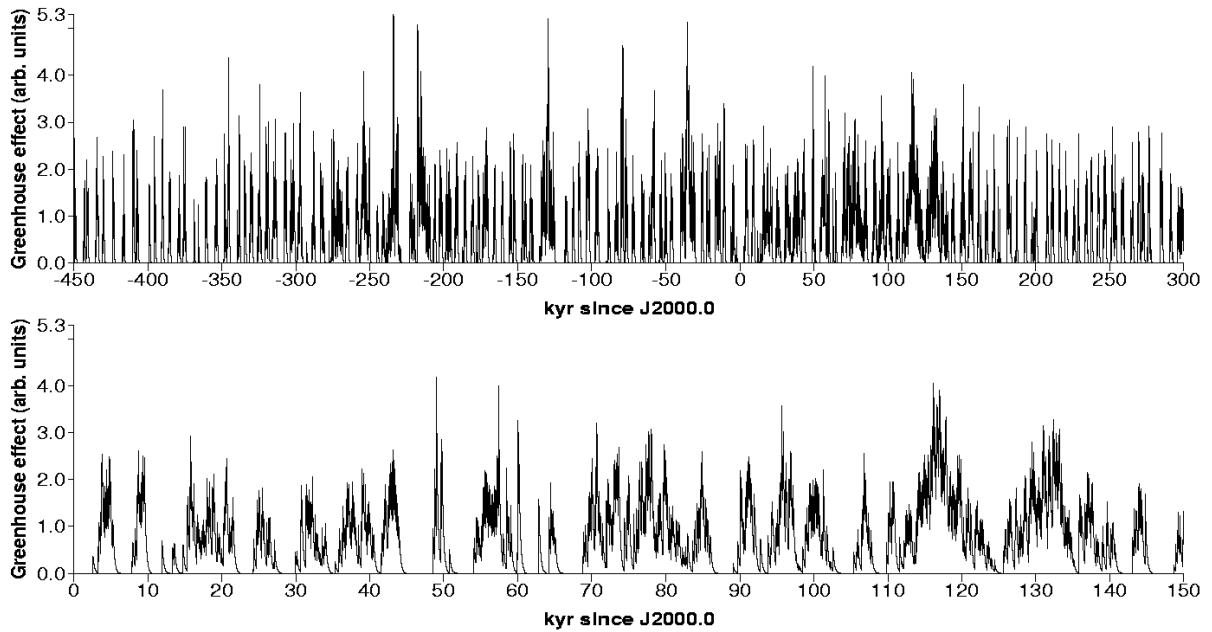}
\caption{Upper panel: Greenhouse spectrum for the distance of
closest approach pattern displayed in Fig.~\ref{MoveZ}. The
vertical scale is in arbitrary units. Lower panel: Expanded view
of this spectrum over 150 kyr shows the similarity between the
calculated structure and that of the Dansgaard-Oeschger events
displayed in Fig.~\ref{IceEnd}.} \label{Greenhouse}
\end{figure*}

\begin{figure*}
\includegraphics[width=16cm,keepaspectratio]{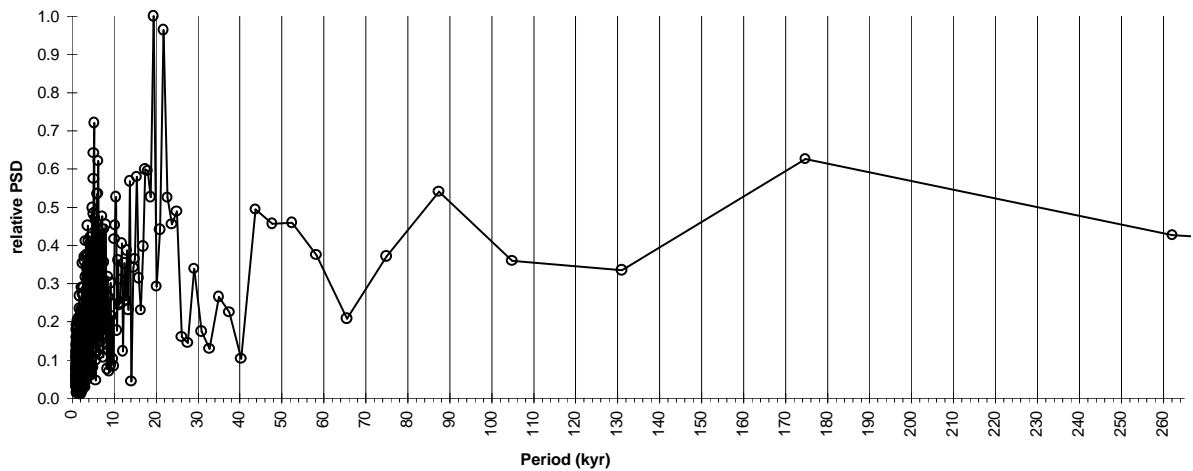}
\caption{Power spectral density (PSD) of the Greenhouse spectrum
presented in Fig.~\ref{Greenhouse}. Visible are three distinct
peaks at 6, 19, 23 kyr, and three broad maxima at ca. 50, 90 and 175 kyr.}
\label{spectrum}
\end{figure*}

Applying Eq.~\ref{11} to the closest approach spectrum given in
Fig.~\ref{MoveZ} yields the Greenhouse heating signal shown in
Fig.~\ref{Greenhouse}. Outgoing and incoming passages were equally
weighted. Actually, incoming gas clouds are more expanded so that
a single passage contributes less. On the other hand they also
contribute for larger distances than those considered in Fig.~\ref{Greenhouse}.
This calculation was performed with a time constant of $\tau_r = 200$
yr which we consider an uppermost limit for the residence time
of the Greenhouse gases of interest here. Again, a one-to-one
correspondence between predicted and observed temperature spikes and
clusters of spikes cannot be expected for the reason given in
chapter 4. Nevertheless, a comparison of the calculated spectrum
with the $\delta^{18}$O polar ice data displayed in Fig.~\ref{IceEnd}
reveals striking similarities in shape, frequency and irregular
clustering. E.g. the spectrum in Fig.~\ref{spectrum} (lower panel)
exhibits 17--26
distinct temperature excursions per 100 kyr, values, which compare
well with the 21 Dansgaard-Oeschger events observed in the polar
ice data (Fig.~\ref{IceEnd}) and some other palaeoclimate archives
during the last glacial period, i.e. 90--11.5 kyr ago \cite{Adkins}.

In order to determine the frequency spectrum of the Greenhouse
signal presented in Fig.~\ref{Greenhouse}, we applied the Fast
Fourier Transformation (FFT) algorithm to $2^{19}$ data points of
this spectrum, i.e. to the time range from +150 to -374 kyr. The
resulting power spectral density as a function of the reciprocal
value of the frequency (the periodicity) is shown in Fig.~\ref{spectrum}.
Distinct maxima can be seen at 6, 19, 23, and ca. 50, 90 and 175 kyr.
These periodicities are the result of the interplay between the
complex movement of Z's and Earth's orbital planes and should not
be confused with similar values delivered by the Milankovitch theory.
Here, the decisive parameters are the oscillation of both inclinations
(about 7 kyr, and 100 kyr, respectively) and the movements of both
perihelia \cite{Nufer}. In the high frequency domain these values
compare quite well with those obtained from comparable analysis of
the GRIP data displayed in Fig.~\ref{IceEnd}. The peak observed at
90 kyr is close to the famous 100 kyr observed in the deep see sediment
data. This suggest that the Interstadials might be the result of an
enhanced clustering of the Dansgaard-Oeschger events. It should be noted,
however, that such a comparison is problematic for two reasons: First
of all, the results of our spectral analysis depend on the selected
orbital parameters and in particular on the mass of Z. Therefore, we
can only say that the evaluated periodicities of the postulated
Greenhouse effect are not in contradiction with observation. Second,
any comparison with deep sea sediment data is questionable, because
the time scale of these measurements is usually tuned to the 100 kyr
cycle of Earth's orbital eccentricity.

\section{Earth's past climate: a record of planet Z's history}
\begin{figure*}
\includegraphics[width=16cm,keepaspectratio]{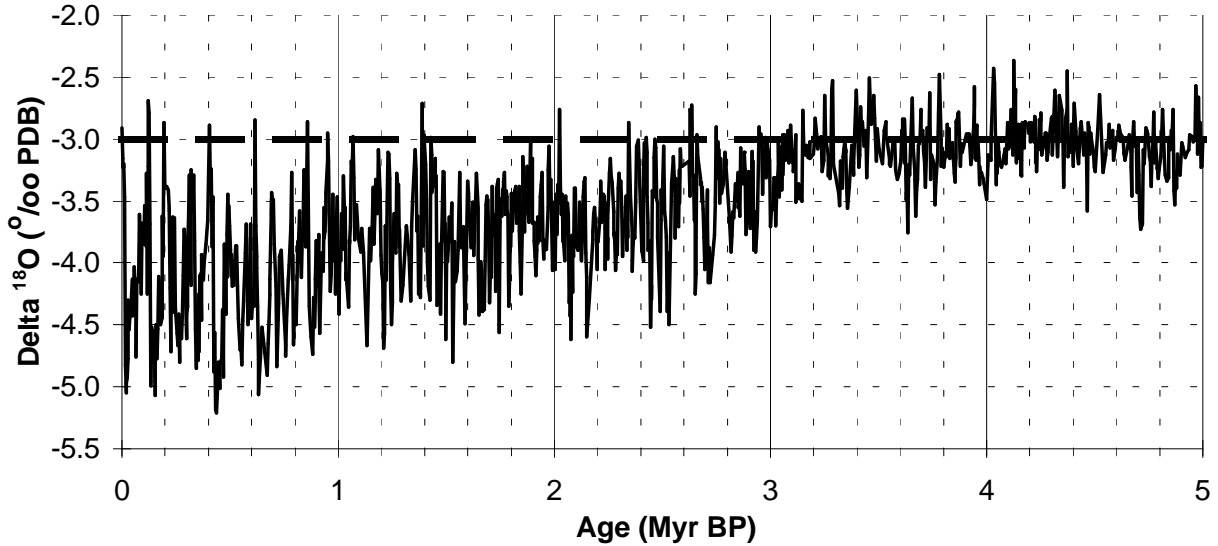}
\caption{Benthic foraminifera $\delta^{18}$0 record from site 659
(18$^\circ$ 05'N, 21$^\circ$ 02'), according to Ref.\protect\cite{Clemens}.
This spectrum is a proxy for the extension and variation of the polar
ice shields over the last 5.0 My. Increasing $\delta^{18}$O corresponds
to decreasing ice caps and to warmer climate. The late Pliocene
(before 3.2 Myr) is characterized by small ice caps and remarkably
stable climatic conditions comparable to those of the Holocene. The
corresponding value is shown by the horizontal dashed line.}
\label{foram}
\end{figure*}

In chapter 1 we pointed to the strange fact that Earth's climate
was remarkably stable prior to the Ice Age epoch. Actually,
the $\delta^{18}$O record of benthic foraminiferas from the
Ocean Drilling program, site 659,
shown in Fig.~\ref{foram}, suggests three different climate
regimes for the last 5.0 Myr \cite{Tiedemann,Clemens}.
The first phase is characterised by a very stable climate with
small polar ice shields, and, in comparison with the following
regimes, high temperature if it is assumed that ice extension
and mean global temperature are correlated. This climate state
is similar to that of the Holocene. A
sudden drop in the global mean temperature terminated this state
at about 3.2 Myr ago. In the following period the variability of
the climate began to increase until about 1.0 Myr ago. At this
time the proxy dates indicate a further drop of the mean temperature
to the lowest known level in Earth history. This marks the beginning
of the Ice Age epoch discussed in the previous chapter. It is obvious,
that the properties of these three, or actually four different
climate epochs, including the Holocene, cannot be explained in terms
of the original Milankovitch theory, unless changes of solar
irradiance or Earth's orbital parameters according to observations
are postulated. However, the introduction of such changes is hard to
justify.

\subsection{Some more polar shifts}
In view of what has been derived so far the spectrum in
Fig.~\ref{foram} can be interpreted in a new way. Prior to about
3.3 Myr ago planet Z was inactive. If it existed, it was not yet
on a highly eccentric orbit, cold, and without gas cloud. The
climate was stable showing only the moderate changes which
actually reflect the undisturbed Milankovitch effect. In agreement
with this theory no 100 kyr-cycle appears up to this time
\cite{Tiedemann}. The fact that the size of the continental ice
caps during this period was comparable to the present day value
indicate that Earth's North pole was then located either in the
Northern part of the Atlantic, in the Pacific Ocean, or at about
the same position as it is today. It should be possible to
evaluate its actual position using information on the climate in
various regions during the Pliocene, and/or from geomagnetic field
measurements. Around 3.2 Myr ago we assume that planet Z shifted
Earth's North pole towards the North American continent in the way
described in chapter 3. As a consequence, the continental ice
shields started to grow and with it Earth's albedo, which reduced
the global mean temperature. Of course, this implies that Z was on
an eccentric Earth crossing orbit not later than at this time.
Tidal heating consumed orbital energy so that its major semi-axis
was slowly reduced. As mentioned in chapter 4, we are unable to
decide in which way its eccentricity changed. The growing
amplitudes of the climatic changes between about 3.2 and 1.0 Myr
suggest a slowly growing gas cloud, which indicates an increasing
orbital eccentricity. The tidal effects increased as well, which
accelerated the heating up of Z. About 1.0 Myr ago, a second polar
shift in direction toward the North American continent to the
position assumed in chapter 3 further lowered the global
temperature. The Ice Age epoch began. At this time the gas cloud
was developed to its full size and the resulting Greenhouse
heating created the temperature variations described in the
previous chapter. A third and final polar shift terminated Earth's
Ice Age epoch 11.5 kyr ago, as discussed in chapter 3. The polar
ice caps readjusted to the new position and, because of the
diminishing size of the continental ice sheets, the global
temperature increased during the early Holocene to its present
value.

The suggestion that Earth's pole was shifted at least three times
during the last 3.2 Myr is compatible with the orbital calculations
presented in chapter 4. They show that Z approached the Earth to
polar shift distances about once in a million years, so that more
than one shift event may be expected during the time range considered
here. The direction of the polar shift depends on the geographic
position of the closest approach. Antipode events excite the same
directional shift, while those at positions differing by 180$^\circ$ in
longitude only, or positions with the same longitude but equal northern,
respectively southern latitudes lead to opposite motions. The earlier
shifts occured with encounters outside the Roche limit, which are
more likely and, in view of the larger mass of Z at early times,
sufficient to produce the required deformation.

\subsection{On the origin and fate of planet Z}
At this point it is tempting to speculate on the origin of Z. We see
three possibilities: First, Z was originally on a hyperbolic
trajectory but, during a close encounter with Jupiter, transferred
to the required highly eccentric bound orbit. The major semi-axis
$A$ of such an orbit would be larger than half of Jupiter's orbital
radius, i.e. $A > 2.6$ AU, because of the presumably large initial
velocity at the inflection point. Such a large $A$ is not compatible
with our orbital calculations in chapter 4, which suggest $A\approx 1$ AU.
Second, Z was originally a moon of Jupiter which was ejected due to
the coupling with the other moons of Jupiter. Such an emission could
result in a rather small initial velocity and accordingly to a highly
eccentric orbit. Again, this is unlikely for the same reason as before.
The last possibility relates to the Asteroid belt problem. It has been
argued that the asteroids circling the sun at a mean distance of about
2.7 AU could be the remains of a collision between two larger bodies,
one originally bound to the sun, the other presumably entering the solar
system on a hyperbolic trajectory. Assuming that Planet Z was a
reaction product of such a collision emerging with low velocity,
then the major semi-axis of its orbit would be not much larger than
about $A = 1.35$ AU at the beginning of its journey around the sun.
Considering the time required to reduce the orbital energy and to
increase its eccentricity to the final values estimated in chapter 4,
it is the most likely scenario.

It is open to discussion at what time Z disappeared. One possibility
is that in the last close encounter it lost its angular momentum and
plunged into the Sun. However, this requires an excentricity of not
less than 0.98, and it supposes that the angular momentum received
is opposed to the existing one. Therefore it seems more likely that
the last encounter with the Earth was within the Roche limit, so that
Z split into several parts. As a result of the reduced escape velocities
of the debris, these parts lost material at an accelerated rate and
were completely dissolved during the Holocene. The gas and dust clouds
were expelled from the planetary system by radiation pressure.

\section{A new understanding of Earth's past climate}

The previous chapters showed that the observed complex structure of
Earth's climate changes during the last few Million years can be
explained in terms of a single object of planetary size, which was bound
to the sun on a highly eccentric orbit for a limited time. This planet,
called Z, was surrounded by a gas cloud and interacted gravitationally
and through its gas cloud with the Earth and its atmosphere with a
frequency determined by the pattern of its closest approaches to the
Earth. These effects influenced Earth's climate in two different ways.

Each polar shift event started with a huge flood, exceptionally strong
winds, earthquakes and volcanic activities and ended with a displaced
pole position after a couple of years. Following this shift the polar
ice caps adjusted to the new pole position which changed Earth's
reflectivity, the albedo, and accordingly the global temperature. The
new pole position determined whether this temperature was higher or
lower after the shift. In agreement with observation we found that this
happened about three times during the last 3.2 Myr. Two times the
temperature dropped, suggesting that each time the pole was shifted
closer towards the North American continent, so that the size of the
continental part of the polar ice cap increased. During the last event,
11.5 kyr ago, it was moved into the arctic sea. At this position a net
reduction of the polar ice caps prevailed. As a result of the reduced
albedo, the global temperature increased to the presently known value.
We also found that Z passed the Earth about 6 times in 100 kyr at
distances less than 300\thinspace 000 km. All of these flyby's
were close enough to excite large tides followed by earth quakes and
volcanic activities. Recently, it has been suggested that the so called
Heinrich events observed in the North Atlantic sediments during the last
100 kyrs --- a rare, but distinct ice rafting phenomenon --- could be the
result of unusual, strong earthquakes \cite{Heinrich,Hunt}. It is
interesting to note that the observed 6 Heinrich-events per 100 kyr
compare well with the predicted frequency for sub-lunar flybys.

The interaction between the gas cloud of Z and Earth's atmosphere was
quite spectacular as well, since most of the kinetic energy of the
particles stopped in the upper atmosphere was converted into radiation.
Probably, this radiation was intense enough to even ignite extended fires
on Earth's surface. After a couple of days the fires were extinct either
by the flood, by heavy rain or by snow enriched with nitrous acid. When
the atmosphere was cleared of the light absorbing components, the
radiative forcing of the produced Greenhouse gases began to prevail,
so that globally the temperature rapidly increased within a few years
to the maximum value determined by the amount of Greenhouse gases produced
during the encounter. The Greenhouse effect was a transient phenomenon,
whose impact was limited by the amount of the Nitrogen compounds
produced per encounter and the residence time of these gases. It may
be responsible for the Dansgaard-Oeschger events observed during the
Late Pleistocene. If so, then the Interstadials may be explained in
terms of the quasi-periodic clustering of these events, as shown in
chapter 5.2.

Summarizing, the many abrupt climate changes characterising Earth's
Ice Age epoch may reflect the interaction between Z and Earth during
the last 3.2 Myr. These interactions were global catastrophes, the worst
of which resulted in a polar wandering. These particular events even
irreversibly changed Earth's climate. The last polar shift event
occurred only 11.5 kyr ago. As pointed out in chapter 2, the violence
of this event is well documented by the observed sudden death of many
animals, the extinction of some of the larger mammal species and the
sudden rise of the sea level \cite{Fairbanks}. Possibly the traditions
in old cultures of catastrophic events refer to the events mentioned
above. Presently, we are enjoying a perfectly Milankovitch-driven
moderate climate without the extreme and fast temperature variations
observed prior to the last polar shift event. This may be due to the
fact that planet Z is no longer influencing the Earth. We are confident
that this new and pleasant climate state, called Holocene, shall last
for an indefinite time, provided, we are willing to stop our own
``Greenhouse attack'' on Earth's climate.

Although we have presented a consistent explanation of the complex
structure of Earth's climate variation during the last few Million years,
we are aware that this rather exotic story has opened up more questions
than it answered, because many of our arguments are based on considerations
which are in terms of order of magnitude and very elementary. Nevertheless,
we believe that this proposal merits further investigation, in view of
its basic simplicity. In particular, we encourage state of the art
calculations which could narrow the allowed parameter range of the
hypothetical planet, or even disprove its existence. Of importance would
be model calculations for the global climate for both pole position, with
and without the suggested additional Greenhouse effect. Our static approach
to Earth's deformation during a fly-by should be replaced by a dynamical
calculation. The actual amount of material that evaporates from a large body
during its closest approach to the sun deserves a more detailed investigation.
The sequence and efficiency of reactions initiated by an energetic atom,
mostly Oxygen and Silicon, impinging on and stopped in the uppermost
atmosphere, has to be further clarified, if necessary, with the help of
appropriate experiments. A determination of the maximum age of the ice at
the Siberian side of the Arctic and/or at the Atlantic side of the Antarctic
could provide a test of the polar shift theory. Planet Z must have left
its chemical signature in polar ice. The material was deposited in atomic
or molecular form, in contrast to dust which is mainly of terresrial origin.
Samples from the Moon could also reveal the presence of Z. These chemical
traces might decide in which way Z disappeared from the solar system.
If the dust cloud which envelops the inner solar system near the ecliptic
plane originated from the last fragments of Z, then the intensity of the
zodiacal light might show a measurable decrease in time~\cite{Gustafson}.
We are optimistic, that the combined results of such studies will finally
converge to a new understanding of Earth's Ice Age epoch. A negative answer
implies that the strange variability of the climate during this epoch, as
well as the stepwise cooling of Earth's climate beginning about 3.2 Mio
years and its abrupt ending only 11.5 kyr ago remain unexplained. We are
sceptical that these problems can be solved by means of any of the presently
favoured Milankovich-based theories alone.
\section*{Acknowledgements}
The authors profited from many discussions, in particular with J\"urg Beer,
Alain Geiger, Hans-J\"urg Gerber, Friedrich Heller, Heinrich Keller, Ernest
Kopp, Roberto Marquardt, Christian Schl\"uchter, Markus Sigrist, Johannes
Staehelin, Martin Suter and Hans-Arno Synal. Our special thanks go to
Hans-Ude Nissen.

\end{document}